\documentclass[a4paper]{jpconf}
\usepackage{graphicx}
\usepackage{graphics}
\usepackage{iopams}
\usepackage[all]{xy}

\newcommand{\half}{\mbox{$\textstyle \frac{1}{2}$}}

\newcommand{\re}{\mbox{$\rm e$}}

\newcommand{\rd}{\mbox{$\rm d$}}

\begin{document}
\title{Constrained quantum dynamics}

\author{Anna~C~T~Gustavsson}

\address{
Blackett Laboratory,
Imperial College London, London SW7 2AZ, UK}

\ead{anna.gustavsson05@imperial.ac.uk}

\begin{abstract}
We give an overview of the two different methods that have been
introduced in order to describe the dynamics of constrained
quantum systems; the symplectic formulation and the metric
formulation. The symplectic method extends the work of Dirac on
constrained classical systems to quantum systems, whereas the
metric approach is purely a quantum mechanical method having no
immediate classical counterpart. Two examples are provided that
illustrate the nonlinear motion induced by the constraint.
\end{abstract}

\section{Introduction}
A question that has recently been asked is how does one implement
unitary motion on a quantum system that is subject to a set of
constraints? In classical mechanics, constrained evolutions are
often constructed using an approach developed by Dirac
\cite{Dirac-50,Dirac-58}, which uses the symplectic geometry of
the classical phase space. Recently, several authors
\cite{Buric,Corichi,BGH-CQM1} have worked on extending the
classical theory of constraint to quantum systems, using the phase
space formulation of quantum mechanics. There are, however, pros
and cons in this approach: It is for example only applicable to
systems subject to an even number of constraints. To address this
and other issues, an alternative approach for treating constrained
quantum systems was introduced in \cite{BGH-CQM2}. In this paper
we review both these approaches and explore through examples the
kind of nonlinear motion that results from constraining a unitary
evolution in quantum mechanics.

The classical method uses the property that the classical phase
space is a symplectic manifold endowed with a symplectic
structure. The quantum mechanical phase space has the geometry of
a K\"ahler manifold, with both a symplectic structure and a metric
structure. As the method considered by
\cite{Buric,Corichi,BGH-CQM1} uses the symplectic geometry of the
quantum phase space, it was given the name the \emph{symplectic
approach} to quantum constraints \cite{BGH-CQM1}. The alternative
formalism introduced in \cite{BGH-CQM2} makes a novel use of the
metric structure of the quantum state space, thus referred to as
the \emph{metric approach} to quantum constraints.

In this paper we will first review some of the main features of
the phase space formulation of quantum mechanics. We will then
highlight the key features of both the symplectic and metric
approaches to implementing constrained unitary motion and discuss
the resulting evolutions by looking at two examples.

\section{The phase-space formulation of quantum mechanics}

The phase space description of quantum mechanics is based on a
projective Hilbert space formulation (see for example
\cite{strocchi,cantoni,Kibble,Weinberg, cirelli, gibbons,
hughston, ashtekar,gqm} and references cited therein). The idea is
as follows. The expectation value of any observable in quantum
mechanics is independent of the norm of the state vector
\begin{eqnarray}
|x\rangle \sim \lambda|x\rangle \qquad \lambda \in \mathbb{C}-\{0\}.
\end{eqnarray}
A state in quantum mechanics can hence be viewed as a ray through
the origin of the Hilbert space $\mathcal H^{n+1}$, and we can,
without loss of any physical information about the system,
consider quantum mechanics in the space of rays. The space of rays
correspond to the complex projective space $\mathbb P^n$, where
each state of the system is now given by a point. In real terms
$\mathbb{P}^n$ is an even dimensional manifold $\Gamma^{2n}$
equipped with a Riemannian structure given by the Fubini-Study
metric $g_{ab}$ and a compatible symplectic structure
$\Omega_{ab}$. The transition probability between two states in
$\Gamma$ is given by the associated geodesic distance between the
two points. It is interesting to note that the probabilistic
aspects of quantum theory are hence associated with the underlying
metric geometry of the state space. In comparison the dynamical
aspects of the theory are captured by underlying symplectic
geometry. Physical observables are functions on $\Gamma$ given by
the expectation value of the corresponding Hilbert space operator
at each point $x$. The Hamiltonian, for example, is given by
\begin{eqnarray}
H(x) = \frac{\langle x| \hat{H} | x \rangle}{\langle x| x
\rangle}. \label{H}
\end{eqnarray}
The eigenstates of the Hamiltonian are given by the fixed points
of $H$:
\begin{eqnarray}
\nabla_a H(x) = 0,
\end{eqnarray}
and the corresponding eigenvalues are the values of $H(x)$ at each
fixed point. Unitary evolution is governed by the Schr\"odinger
equation, which takes the form of Hamilton's equations:
\begin{eqnarray} \label{SE}
\frac{\rd x^a}{\rd t} = \Omega^{ab}\nabla_b H(x),
\end{eqnarray}
where $\Omega^{ab}$ is the inverse symplectic structure given by
$\Omega_{ab}\Omega^{bc}=-\delta_a^c$. Here, and throughout the
paper, there is an implied summation over repeated indices. We
parameterise the state space by the ``action-angle''
parameterisation (cf. \cite{oh}) so that the equations of motion
(\ref{SE}) take a particularly simple and familiar form. If we let
the state $|x\rangle$ in energy basis be given by
\begin{eqnarray}
|x\rangle = \sum_{i=1}^n \sqrt{p_i}\re^{-{\rm i} q_i}|E_i\rangle +
\sqrt{1-\sum_{i=1}^n p_i}|E_{n+1}\rangle,
\end{eqnarray}
then the expectation of the Hamiltonian operator $\hat{H}=\sum_j
E_j|E_j\rangle\langle E_j|$ is given by
\begin{eqnarray}
H(q_i,p_i) = E_{n+1} + \sum_{i=1}^n \omega_i p_i, \quad
\textrm{where} \quad \omega_i = E_i - E_{n+1}.
\end{eqnarray}
The equations of motion then take the form
\begin{eqnarray}
\dot{q}_i = \frac{\partial H(q_i,p_i)}{\partial p_i} \quad
\textrm{and} \quad \dot{p}_i = - \frac{\partial
H(q_i,p_i)}{\partial q_i},
\end{eqnarray}
with solutions $q_i(t)=q_i(0) + \omega_i t$ and $p_i(t) = p_i(0)$.

\section{Constrained dynamics}

Let us now assume that the evolution of the quantum system is
subject to one or more constraints. The type of constraints that
we will be considering can be written in the form
\begin{eqnarray} \label{phi}
\Phi^{i}(x)=0, \qquad i=1,2,\ldots,N,
\end{eqnarray}
where $N$ is the total number of constraints.
\begin{figure}[th]
\begin{center}
\includegraphics[width=15pc]{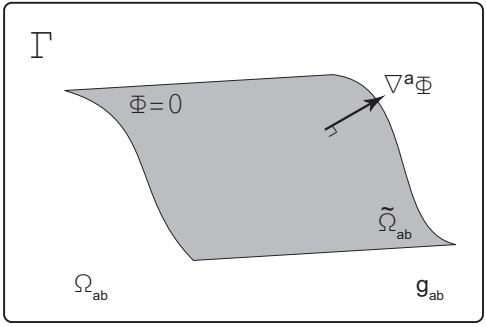}\hspace{1pc}%
\begin{minipage}[b]{20pc}
\caption{The constraints $\Phi^i=0$ define a subspace of $\Gamma$.
The evolution of the system can be constrained to this subspace
using either the symplectic or the metric method. When using the
symplectic approach the modified equations of motion take the same
form as the original one, but with a modified symplectic structure
$\tilde{\Omega}_{ab}$. In the metric approach the system is
constrained by removing all components orthogonal to the surface
from the vector field $\dot{x}^a=\Omega^{ab}\nabla_b H$. \label{fig1}}
\end{minipage}
\end{center}
\end{figure}
Equation (\ref{phi}) defines a subspace $\Phi$ of $\Gamma$, given
by the intersection of each of the $N$ subspaces defined by
$\Phi^i=0$, onto which the evolution will be constrained, see
figure \ref{fig1}.

\begin{figure}[th]
\begin{center}
\includegraphics[width=15pc]{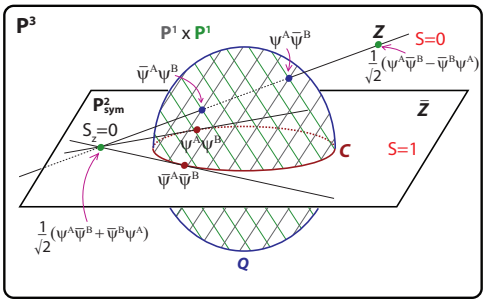}\hspace{1pc}%
\begin{minipage}[b]{20pc}
\caption{The state space of 2 spin-$\half$ particles. The
disentangled states form a subspace of the total state space given
by $Q=\mathbb{P}^1 \times \mathbb{P}^1$, where $\mathbb{P}^1$ is
the complex projective line. Further details of the state space
structure of this system can be found in \cite{gqm,bgh1}.
\label{fig2}}
\end{minipage}
\end{center}
\end{figure}

We shall here consider two different types of constraints. The
first type is given by the conservation of a set of observables,
i.e.
\begin{eqnarray} \label{ExpPhi}
\Phi^{i}(x)=\frac{\langle x|\hat{\Phi}^{i}|x\rangle}{\langle x|x\rangle},
\end{eqnarray}
where $\hat{\Phi}^{i}$ is the Hilbert space operator corresponding
to the observable $\Phi^i$. The second type of constraint is when
the evolution of the system is constrained to an algebraically
well defined subspace of $\Gamma$ that does not correspond to the
conservation of any observable. An example of this is to constrain
the motion of a pair of spin-$\half$ particles that are initially
disentangled to remain disentangled. There is no observable that
corresponds to this property of the system, but as disentangled
states form a subspace $Q=\mathbb{P}^1\times\mathbb{P}^1$
\cite{gqm,bgh1} of the overall state space $\mathbb{P}^3$ (see
figure \ref{fig2}), we can constrain the system algebraically to
remain on this subspace.
We shall return to this example and look at it in more detail in
the next section where we give an outline of the symplectic
approach to implementing quantum constraints.

\subsection{Symplectic approach}

Let us first consider the symplectic approach and follow the
methodology introduced in \cite{BGH-CQM1}. In order to implement
the constraints (\ref{phi}) the equations of motion are modified
such that
\begin{eqnarray} \label{SympSE}
\frac{\rd x^a}{\rd t}=\Omega^{ab}\nabla_b H(x) +
\lambda_{i}\Omega^{ab}\nabla_b\Phi^{i}(x),
\end{eqnarray}
where $\lambda_i$ are Lagrange multipliers associated with the
constraints $\Phi^i(x)$. In some cases these Lagrange multipliers
can be found explicitly by considering $\dot{\Phi}^{i}=0$. By
expanding $\dot{\Phi}^{i}$ using the chain rule $\dot{\Phi}^{i} =
\dot{x}^a \nabla_a \Phi^{i}$ we find that the Lagrange multiplier
is given by:
\begin{eqnarray}
\lambda_{i} = \omega_{ji} \Omega^{ab} \nabla_a
\Phi^{j} \nabla_b H,
\end{eqnarray}
where $\omega_{ij}$, assuming it exists, is the inverse of
\begin{eqnarray} \label{omega}
\omega^{ij}:=\Omega^{ab} \nabla_a \Phi^{i} \nabla_b \Phi^{j}.
\end{eqnarray}
Note that when the constraints are given by conserved observables
(\ref{ExpPhi}), then $\omega^{ij}$ correspond to the commutator
between the two observables. By substituting the Lagrange
multipliers back into (\ref{SympSE}) we find that the equations of
motion can be rewritten in the form \cite{BGH-CQM1}
\begin{eqnarray} \label{EOMwithOmega-tilde}
\frac{\rd x^a}{\rd t}=\tilde{\Omega}^{ab}\nabla_b H(x),
\end{eqnarray}
where $\tilde{\Omega}^{ab}$ is the induced symplectic structure on
the constraint subspace $\Phi^i=0$ given by
\begin{eqnarray} \label{omega-tilde}
\tilde{\Omega}^{ab}=\Omega^{ab} + \Omega^{ac}\Omega^{bd}
\omega_{ij} \nabla_c\Phi^{i}\nabla_d \Phi^{j}.
\end{eqnarray}
In other words, the equations of motion for the constrained system
take the same form as for the unconstrained system. Only the
symplectic structure is no longer the globally defined
$\Omega_{ab}$, but the symplectic structure induced locally on the
subspace $\Phi^i=0$ (see figure \ref{fig1}). A consequence of this
is that we can no longer expect the resulting motion to be
unitary, as the induced symplectic structure $\tilde{\Omega}_{ab}$
has a nonlinear term. In order to illustrate the kind of motion
that could result from this kind of constraint we consider the
example of two spin-$\half$ particles constrained to remain
disentangled, as mentioned above. Let us first note however that
the symplectic approach is not applicable to all systems. In order
to use it, one requirement is that $N$, the total number of
constraints being imposed, must be even. This follows by
construction as otherwise the symplectic geometry is not preserved
on the constraint surface. Another condition that must be
satisfied in order to find the lagrange multipliers is that the
inverse of (\ref{omega}) must exist.

\textbf{Example 1:} Let us now consider the example of two
spin-$\half$ particles with Hamiltonian:
\begin{eqnarray}
{\hat H}=-J{\hat{\sigma}}_1\otimes {\hat{\sigma}}_2 -B
(\hat{\sigma}^z_{1} \otimes{\mathbf 1}_2 + {\mathbf 1}_1 \otimes
\hat{\sigma}^z_{2}),
\end{eqnarray}
where the subscripts 1 and 2 labels the two particles, $J$ is the
strength of the spin-interaction, $B$ is the strength of the
external magnetic field (orientated along the $z$-axis), and
$\hat{\sigma}_z$ is the Pauli spin matrix in the $z$-direction. We
further require the initial state to be disentangled, and we wish
to constrain the system so that it remains so. In other words we
want the state of the system to remain on the surface
$Q=\mathbb{P}^1\times\mathbb{P}^1$ as it evolves. To find the
equations of motion we follow the method outlined above, first
finding (\ref{omega}), taking its inverse, using it to evaluate
the Lagrange multipliers and hence finding $\tilde{\Omega}_{ab}$
(please see \cite{BGH-CQM1} for full details of this calculation).

\begin{figure}
\begin{center}
\begin{minipage}[l]{0.49\textwidth}
\begin{center}
\includegraphics[width=0.95\textwidth]{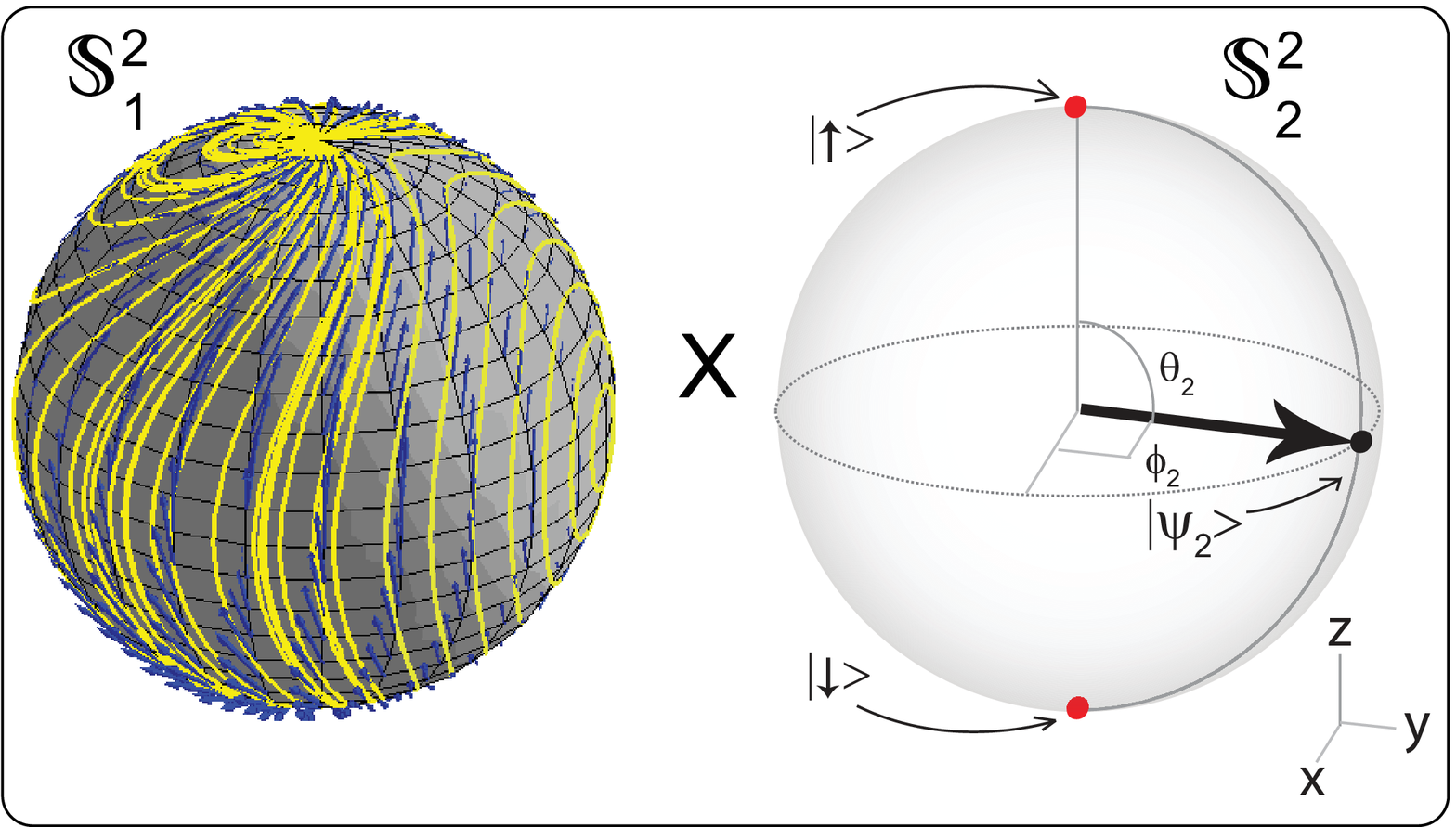}
\end{center}
\end{minipage}
\begin{minipage}[r]{0.49\textwidth}
\begin{center}
\includegraphics[width=0.95\textwidth]{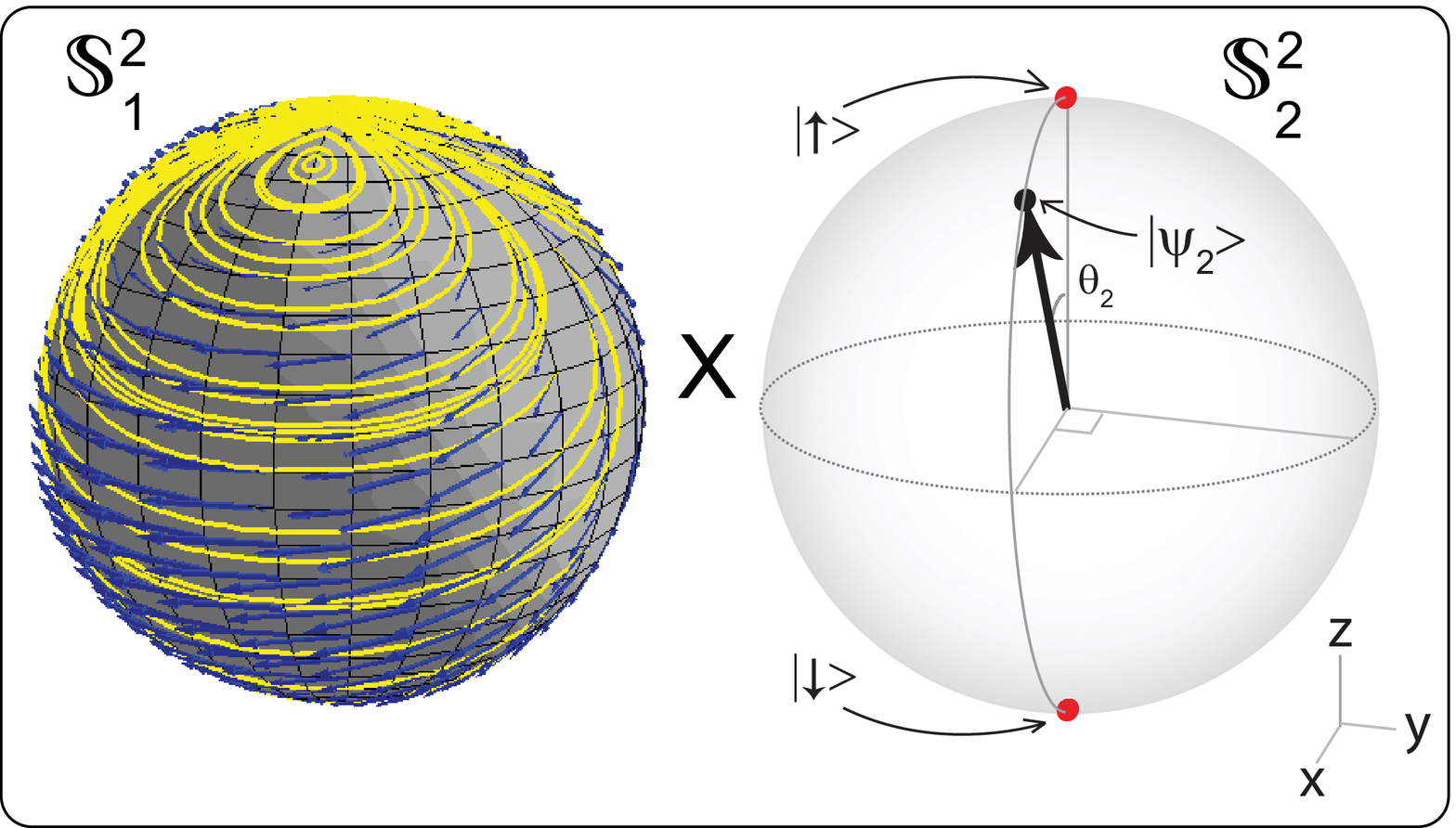}
\end{center}
\end{minipage}
\begin{minipage}[l]{0.49\textwidth}
\begin{center}
\includegraphics[width=0.95\textwidth]{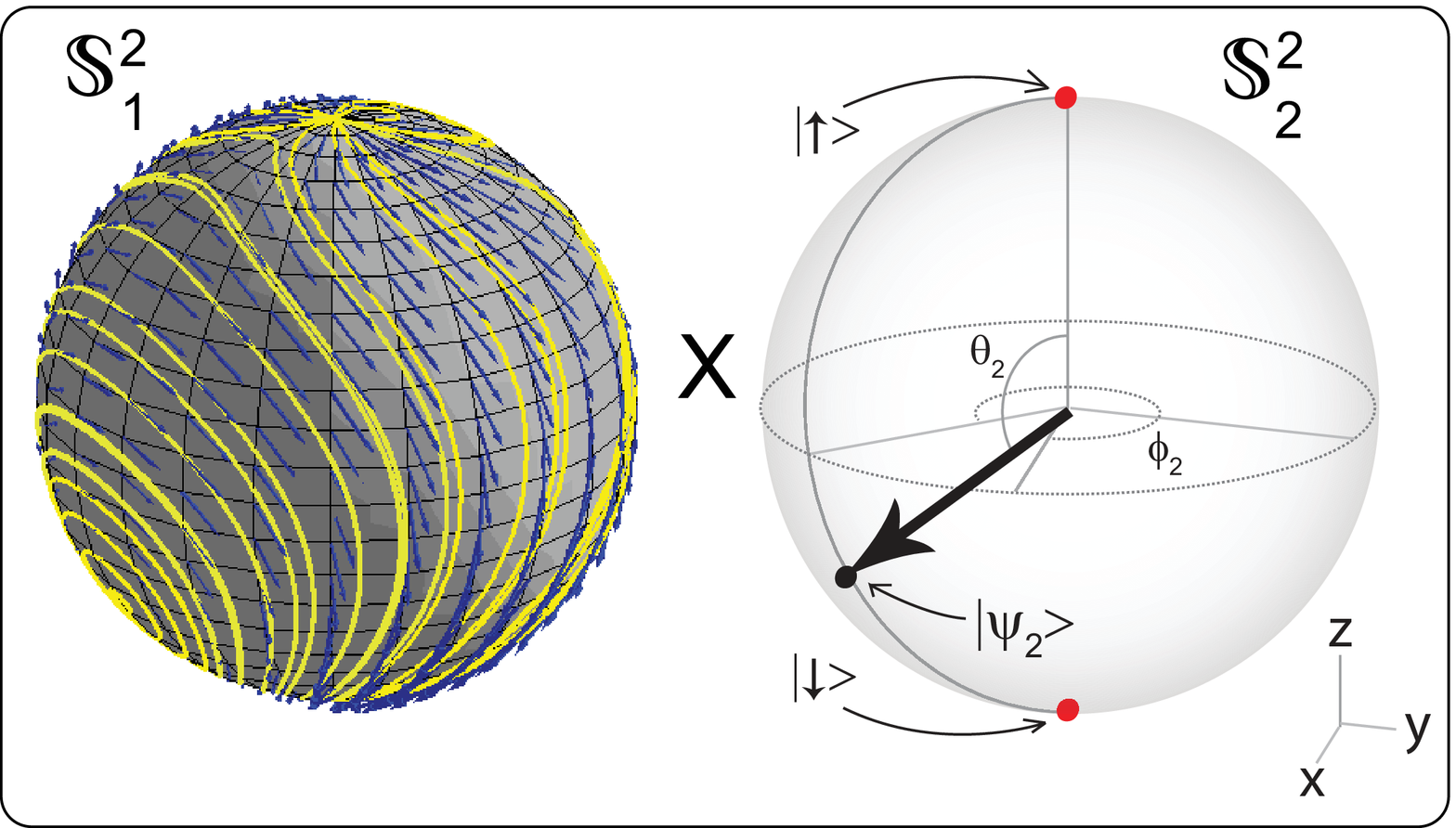}
\end{center}
\end{minipage}
\begin{minipage}[r]{0.49\textwidth}
\begin{center}
\includegraphics[width=0.95\textwidth]{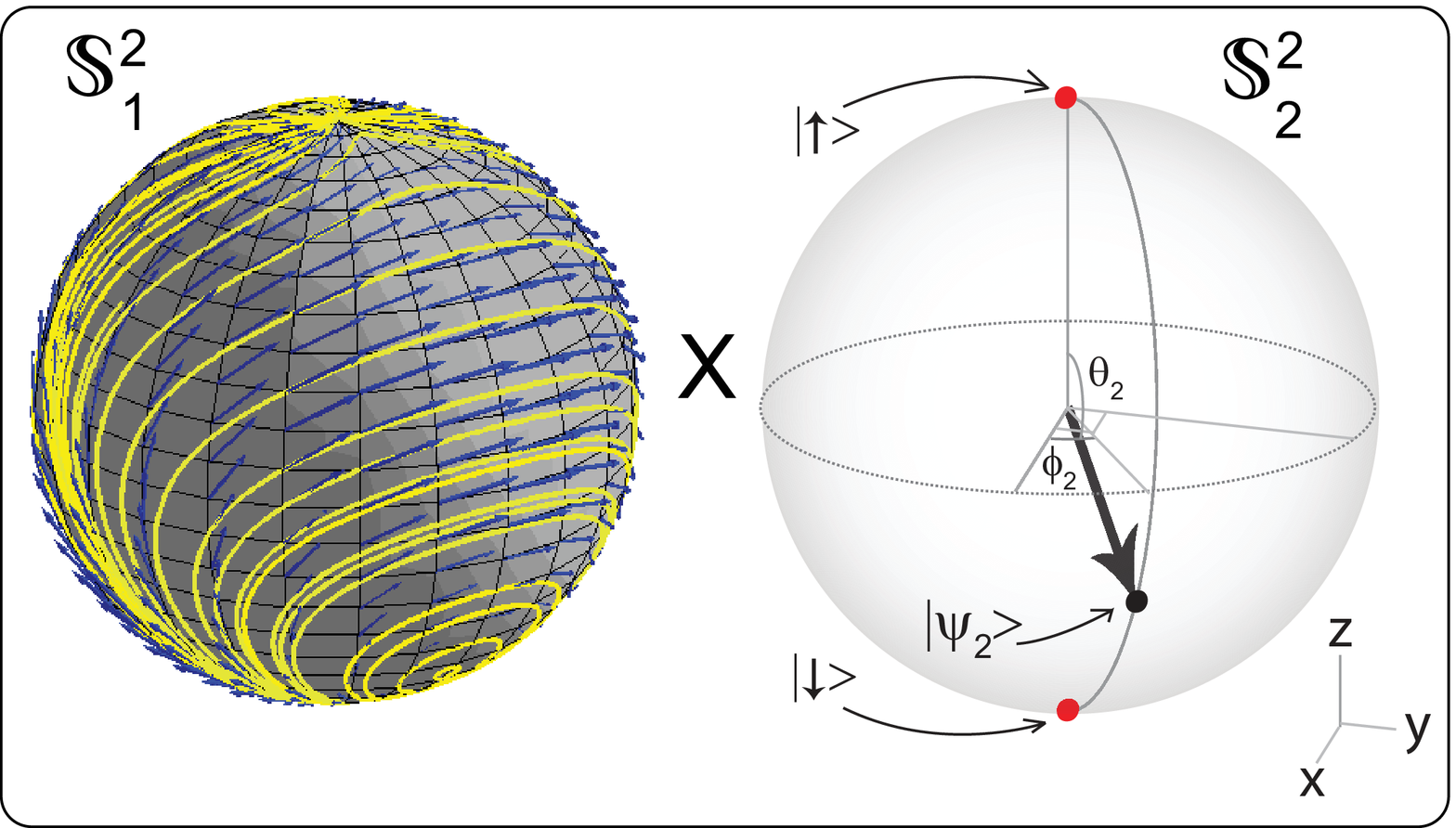}
\end{center}
\end{minipage}
\caption{\label{fig3} Four examples of `snapshots' of the vector field and its integral curves as two spin-$\half$ particles are constrained to remain disentangled so that their evolution lies on the product space $\mathbf{S}^2_1 \times \mathbf{S}^2_2$. The figures show the vector fields on $\mathbf{S}^2_1$ as the second particle evolves through the states $(\theta_2=\frac{\pi}{2}, \phi_2=\frac{\pi}{2})$ (top left), $(\theta_2=\frac{\pi}{6}, \phi_2=0)$ (top right), $(\theta_2=\frac{2\pi}{3}, \phi_2=\frac{5\pi}{3})$ (bottom left) and $(\theta_2=\frac{3\pi}{4}, \phi_2=\frac{\pi}{4})$ (bottom right), on $\mathbf{S}^2_2$.}
\end{center}
\end{figure}

Since the complex projective line is isomorphic to the two-sphere
$\mathbb{P}^1\sim\mathbb{S}^2$, we can view $Q$ as the product
space of a pair of 2-spheres $\mathbb{S}_1^2\times\mathbb{S}_2^2$.
Changing coordinates into spherical coordinates (see
\cite{BGH-CQM1}) we can visualise the evolution on the surface of
the 2-spheres. The resulting equations of motion consist of four
coupled non-linear differential equations:
\begin{eqnarray}
\begin{array}{l}
\dot{\theta}_1 = \sin(\phi_1 - \phi_2)
\sin\theta_2 [ (\omega_1 - \omega_2) \cos\theta_1 + \omega_2 -
\omega_3]\\
\dot{\theta}_2  =  \sin(\phi_1 - \phi_2) \sin\theta_1
[(\omega_2 - \omega_1)\cos\theta_2 - \omega_2 + \omega_3] \\
\dot{\phi}_1 =  \frac{1}{2} [-\omega_1 + (\omega_2 -
\frac{\omega_1}{2})\cos\theta_2 + (\frac{3}{2} \omega_1 - \omega_2
-2\omega_3)\cos\theta_1 + [\cos(\phi_1-\phi_2)/(\sin \theta_1
\sin\theta_2)] \\ \qquad \quad \times \left.\left(2
(\omega_3-\omega_2) \sin^2\theta_1 \cos\theta_2 + (\omega_1-
\omega_2) (\cos^2\theta_1 - \cos^2\theta_2) \right)\right] \\
\dot{\phi}_2  =  \frac{1}{2} [-\omega_1 + (\omega_2 -
\frac{\omega_1}{2})\cos\theta_1 + (\frac{3}{2} \omega_1 - \omega_2
-2\omega_3)\cos\theta_2 + [\cos(\phi_1-\phi_2)/(\sin\theta_1
\sin\theta_2)] \\ \qquad \quad \times \left.
\left(2(\omega_3-\omega_2)\cos\theta_1 \sin^2\theta_2 - (\omega_1
-\omega_2)(\cos^2\theta_1 - \cos^2\theta_2) \right) \right].
\end{array}
\end{eqnarray}
It is difficult to solve these equations of motion in closed form.
We will therefore explore the solution numerically by looking at
various `snapshots' of the vector field and associated
trajectories as the system passes through certain points. In
figure~\ref{fig3} we see four examples of this. The top left
picture shows us the vector field on $\mathbb{S}^2_1$ as the
evolution on $\mathbb{S}^2_2$ passes through the point $\theta_2 =
\phi_2 = \half\pi$. As the overall state of the system changes, we
evolve on both spheres and as we change state on $\mathbb{S}^2_1$
the vector field on $\mathbb{S}^2_2$ changes and vice versa. To
fully explore the solution numerically, we require an `interactive
simulation'; figure~\ref{fig3} merely gives a flavour for the
often complicated trajectories that result from constraining the
system to remain on $Q$. The strength of the formalism is that the
solutions are nonetheless fully tractable, at least numerically.

\subsection{Metric approach}
In the previous section we found that there is a way to implement
quantum constraints using the underlying symplectic geometry of
the phase space. Let us now look at how some of the limitations of
the symplectic formalism can be overcome by using the metric
geometry of the phase space.

Another way of constraining the evolution of a system whose
initial state lies on the surface $\Phi=0$ is to remove all the
components of the vector field normal to it \cite{BGH-CQM2} (see
figure~\ref{fig1}). This will force the system to remain on
$\Phi=0$. To implement this constraint we still use the method of
Lagrange multipliers, but we substitute the symplectic structure
in the second term of (\ref{SympSE}) with the metric structure.
Hence the modified equations of motion become:
\begin{eqnarray} \label{SEmetric}
\frac{\rd x^a}{\rd t}=\Omega^{ab}\nabla_b H - \lambda_{i}g^{ab}\nabla_b\Phi^{i},
\end{eqnarray}
where $g^{ab}$ is the inverse of the Fubini-Study metric. Because
we are now making use of the metric geometry of the constraint
subspace, we no longer require that the total number of
constraints $N$ be even. The Lagrange multipliers $\lambda_{i}$
can again be found explicitly in a similar way to the previous
approach by considering $\dot{\Phi}^{i}=0$. Following the method
in \cite{BGH-CQM2}, we hence find that
\begin{eqnarray} \label{metric-lambda}
\lambda_{i} = M_{ij} \Omega^{ab} \nabla_a
\Phi^{j} \nabla_b H,
\end{eqnarray}
where $M_{ij}$ is the inverse of the matrix
\begin{eqnarray}\label{M}
M^{ij}:=g^{ab} \nabla_a \Phi^{i} \nabla_b \Phi^{j}.
\end{eqnarray}
For the case when the constraints correspond to a family of
conserved quantum observables, the matrix $M^{ij}$ corresponds to
the anticommutators between the observables. We hence require
${\rm det}(M^{ij}) \neq 0$ in order to find the equations of
motion:
\begin{eqnarray}\label{metric-eom}
\dot{x}^a = \Omega^{ab} \nabla_b H - M_{ij}
\Omega^{cd}\nabla_c \Phi^{j} \nabla_d H g^{ab} \nabla_b
\Phi^{i}.
\end{eqnarray}
It is no longer in general the case that these can be rewritten in
the form $\dot{x}^a=\tilde{\Omega}^{ab}\nabla_b H$, where
$\tilde{\Omega}^{ab}$ is the inverse of a modified symplectic
structure. We would, however, expect that systems that can be
treated with both methods return the same result. It turns out
that this is very hard to verify in general. In \cite{BGH-CQM2}, a
sufficient condition was derived under which the metric approach
reduces to the symplectic approach with the modified equations of
motion taking the form of (\ref{EOMwithOmega-tilde}), (see
\cite{BGH-CQM2} for further details on this). We therefore take
the view that the metric approach is the more general way of
treating quantum constraints, but under certain conditions it
reduces to an extension of the classical method to quantum systems
as in the symplectic approach. We shall now look at one of the
simplest examples of a system subject to only one constraint.

\begin{figure}
\begin{center}
\includegraphics[width=0.5\textwidth]{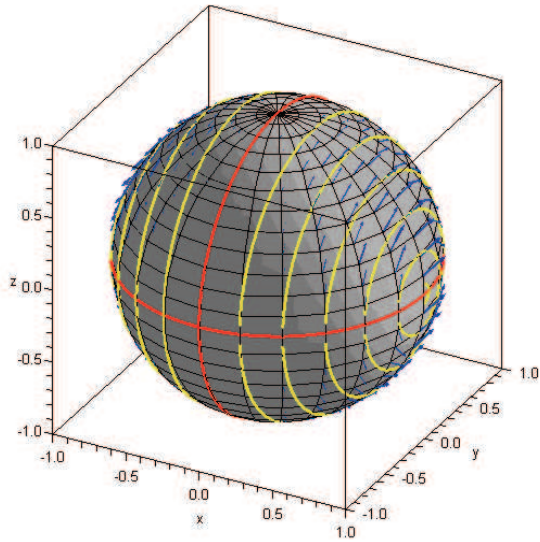}
\caption{The vector field resulting from a single spin-$\half$
particle with Hamiltonian $\hat{H}=\hat{\sigma}_z$ being
constrained such that $A(x)=\frac{\langle
x|\hat{\sigma}_x|x\rangle}{\langle x|x\rangle}$ is conserved
\cite{BGH-CQM2}. The yellow lines are integral curves of the
vector field and the red lines correspond to fixed points.}
\label{fig4}
\end{center}
\end{figure}

\textbf{Example 2:} Let us consider the example of a single
spin-$\half$ particle with Hamiltonian $\hat{H}=\hat{\sigma}_z$,
where $\hat{\sigma}_z$ is a Pauli spin matrix, constrained such
that
\begin{eqnarray} \label{constEx2}
\Phi(x)=\frac{\langle x|\hat{\sigma}_x|x\rangle}{\langle x|x\rangle}
\end{eqnarray}
is conserved \cite{BGH-CQM2}. This example is interesting as we
have an observable that does not commute with the Hamiltonian.
Following the method outlined above, the equations of motion are
obtained by considering (\ref{M}), taking its inverse, finding the
Lagrange multipliers (\ref{metric-lambda}) and hence the equations
of motion (\ref{metric-eom}). Just as in the previous example we
look at the problem in spherical coordinates to more easily
visualise the resulting evolution as trajectories on the surface
of the 2-sphere. The modified equations of motion are given by
\cite{BGH-CQM2}
\begin{eqnarray}
\dot{\theta} = \half \left(\frac{ \sin(2\theta)\sin(2\phi)}
{1-\sin^2\theta\cos^2\phi}\right), \quad {\rm and,} \quad
\dot{\phi} = \frac{2\cos^2 \theta\cos^2 \phi}{1-\sin^2 \theta
\cos^2 \phi}. \label{ex2eom2}
\end{eqnarray}
Figure~\ref{fig4} shows the resulting vector field and some of its
integral curves. We recall that unitary unconstrained motion for
this system corresponds to rigid rotation around the $z$-axis. As
we are now forcing the system to remain on $\langle{\hat
\sigma}_x\rangle=0$, corresponding to circles around the $x$-axis,
the resulting solution has a very interesting fixed point
structure. Instead of only having two fixed points, the poles
along the $z$-axis, as in the unconstrained case, the state space
is now effectively partitioned into quarters of the sphere that do
not interact with each other. Systems whose initial state is not a
fixed point hence evolve along the half circular trajectories
towards the fixed points along the equator.


\ack ACTG would like to thank D C Brody, L P Hughston and D W Hook
for many useful discussions, and for useful comments on an earlier version of this article. ACTG
would also like to thank the organisers of the DICE2008 conference
in Castiglioncello, Italy, 22-26 September 2008, for the
opportunity to present this work there. \vskip10pt


%

\end{document}